\documentstyle[procl]{article}

\input{psfig.sty}

\def\Journal#1#2#3#4{(#1) {#2} {\bf #3}, #4}

\def\AAp{\em Astron. Astrophys.}

\def\ApJ{\em Astrophys.~J.}
\def\ApJL{\em Astrophys.~J., Lett.}

\def\MNRAS{\em Mon. Not. R.~Astron. Soc.}

\begin{document}

\markboth{F. Primini et al.}{Chandra Observations of M31 and their 
				Implications for its ISM}

\thispagestyle{plain}

\title{Chandra Observations of M31 and their Implications for its ISM}

\author{F. Primini, M. Garcia, S. Murray, W. Forman, C. Jones, J. McClintock}

\address{Harvard-Smithsonian Center for Astrophysics,\\
60 Garden Street, Cambridge, MA 02138, USA}

\maketitle

\abstract{We have been regularly observing the bulge and inner disk of
M31 for nearly 1 year, using both the HRC-I and ACIS-I instruments on
the Chandra X-Ray Observatory. We
present results from our program that are of interest to the study of
the ISM in M31. In particular, we find that the unresolved emission
within 3$^{\prime}$ of the center of M31 has a distinctly softer
spectrum than that of most of the resolved x-ray sources in the
region. 
Preliminary spectral analysis of bright point
sources in the bulge shows no evidence (within the poor
statistics) for soft spectral components, but does reveal significant
extra-galactic 
x-ray extinction ($N_{H}\sim 10^{21}$ cm$^{-2}$).
We find no new x-ray counterparts to supernova remnants to date.
}

\section{Introduction}

X-ray Astronomers have long been interested in M31 because
of the opportunity
it affords to study x-ray emission in a galaxy like ours from an
external vantage point (for a current review of the  x-ray properties
of M31, see the paper by M. Ehle in these proceedings ).
Although the emission is
dominated by the contribution from point sources, its study can still
add to our understanding of the ISM in M31.
For example, recent
ROSAT HRI and PSPC observations (Primini, Forman, \& Jones 1993; West,
Barber, \& Folgheraiter 1997) have indicated the presence of diffuse
emission in the bulge and disk. These results are controversial,
however, particularly with respect to the bulge where a
superposition of faint LMXBs with
soft spectral components might appear as diffuse emission (Irwin \&
Bregman 1999; 
Borozdin \& Priedhorsky 2000). Resolution of this issue is
clearly important for the understanding of both the ISM in M31 and the
nature of LMXBs. The Chandra X-ray Observatory, with its
unprecedented combined spectral/spatial capabilities, should provide
the answer. 
In addition, we can use individual x-ray sources to probe the ISM. 
The high-quality spectra available with Chandra
should allow us to accurately determine the  x-ray absorption toward the
brighter sources in the bulge. If we can identify  the
local source contribution, we may be able to provide an
independent estimate of extinction at specific locations in the
ISM. Similarly, by measuring the x-ray surface brightness of supernova remnants
with known diameters  we can constrain the ambient
density in the vicinity of the remnant for typical SNR evolutionary
models(see, e.g., Magnier et al. 1997). 

As part of the Chandra HRC Instrument Principal Investigator
Guaranteed Time Program, we are  monitoring
the inner disk of M31 with the HRC-I and ACIS-I
instruments. 
At monthly intervals for a
year, we are making a series of short ($\sim$1 ksec), overlapping HRC
observations of a $\sim 2^{\circ}$ wide region along the major axis of
M31, centered on the nucleus (see Fig. 1). 
The objective is
to discover x-ray 
transients, which would then be observed in follow-up $\sim$5 ksec
ACIS observations, pre-programmed into the observing schedule at
monthly intervals. Should no transients be detected, each ACIS
observation will be centered on the nucleus.
For the HRC, we expect a point source sensitivity of
$\sim2\times10^{37}$ergs~s$^{-1}$ per month, or
$10^{36}$ergs~s$^{-1}$ after one year (unabsorbed luminosity, assuming
typical x-ray source spectra and $N_H\sim10^{21}$~cm$^{-2}$). 
For ACIS, assuming all observations are centered on the nucleus, the
sensitivities are  
$\sim8\times10^{35}$ergs~s$^{-1}$ per month and
$\sim7\times10^{34}$ergs~s$^{-1}$ per year.
For extended sources, the sensitivities are a factor of $\sim2-10$
worse.

In this paper we present preliminary results
from the first three months of
ACIS observations (prior to the change in CCD operating temperature)
and the first 5 months of HRC observations.
We find that the spectrum of the unresolved emission is softer than
that of most point sources.
No evidence for soft spectral
components is present in individual source spectra, although
it is not
precluded by the limited statistics.
Our search for x-ray-emitting SNRs has to date
yielded no additional candidates beyond those reported by Magnier et
al. (1997) and Supper et al. (1997).

\begin{figure}[htb]
\centerline{\psfig{file=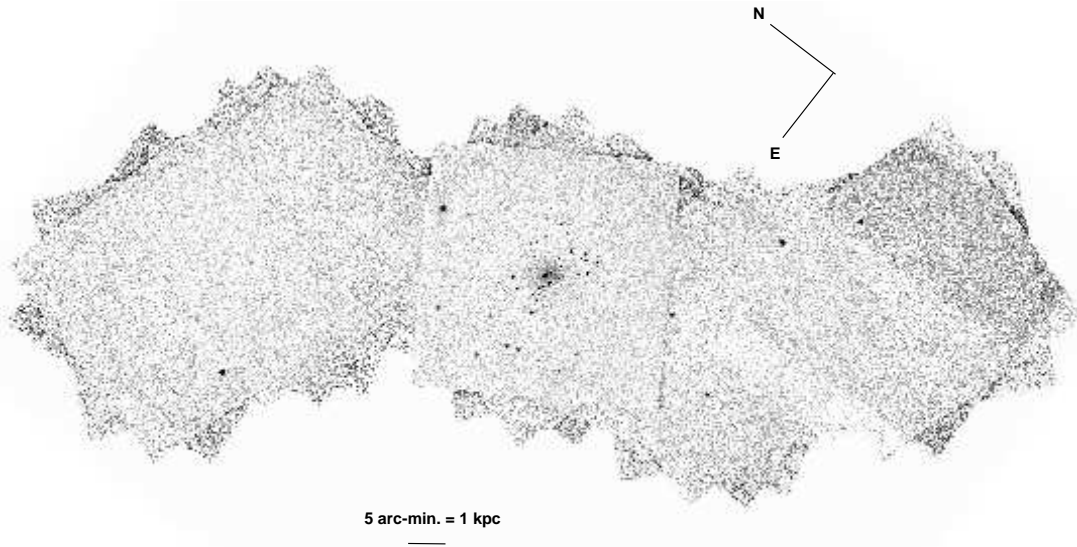,width=145truemm}}
\vspace*{-10truemm}\caption{Chandra HRC Survey of M31. The central field is
$\sim30^{\prime}\times30^{\prime}$ centered on the nucleus, and has an
exposure of $\sim$10 ksec. Exposures for outlying images range are
$\sim4$ ksec., increasing to $\sim$8 ksec. in overlap regions.}
\end{figure}

\section{Unresolved Emission in the Bulge}

ACIS-I images of the central $\sim10^{\prime}\times10^{\prime}$ of M31
are shown in Fig. 2, in two energy bands, from 0.1 - 1.0 keV, and from
1.0 - 7.0 keV. Despite Chandra's excellent imaging capability, significant
x-ray emission in the bulge remains unresolved.
Radial profiles of the emission and hardness ratio, with
contributions from point sources removed, 
are shown in Fig. 3a. 
For comparison, a histogram of the same hardness ratio for
79 point sources in the bulge examined by Garcia et al. (2000) is
shown in Fig. 3b. 
The spectrum of the unresolved emission in the inner 3$^{\prime}$ is
distinctly softer than all but 3 of these sources, one of which is the
nuclear source. For a
bremsstrahlung spectrum with  kT$\sim1$~keV and $N_H\sim10^{21}$~cm$^{-2}$,
the luminosity is $\sim3\times10^{38}$~ergs~s$^{-1}$.
These results support our earlier claim (Primini, Forman, \& Jones
1993) that the emission is either diffuse or a superposition of emission from 
a class of sources different from the bright bulge sources.

\begin{figure}[htb]
\centerline{\psfig{file=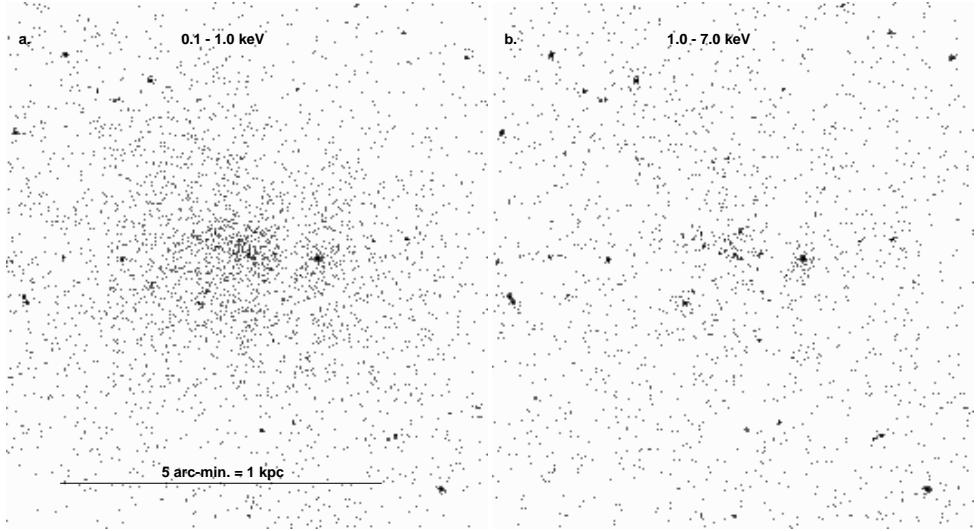,width=130truemm}}
\caption{Chandra ACIS-I Image of the Bulge of M31. Total exposure is
$\sim$15 ksec. The nearly vertical stripe near the center is due to a
gap in the ACIS chips.}
\end{figure}

\begin{figure}[htb]
\centerline{\psfig{file=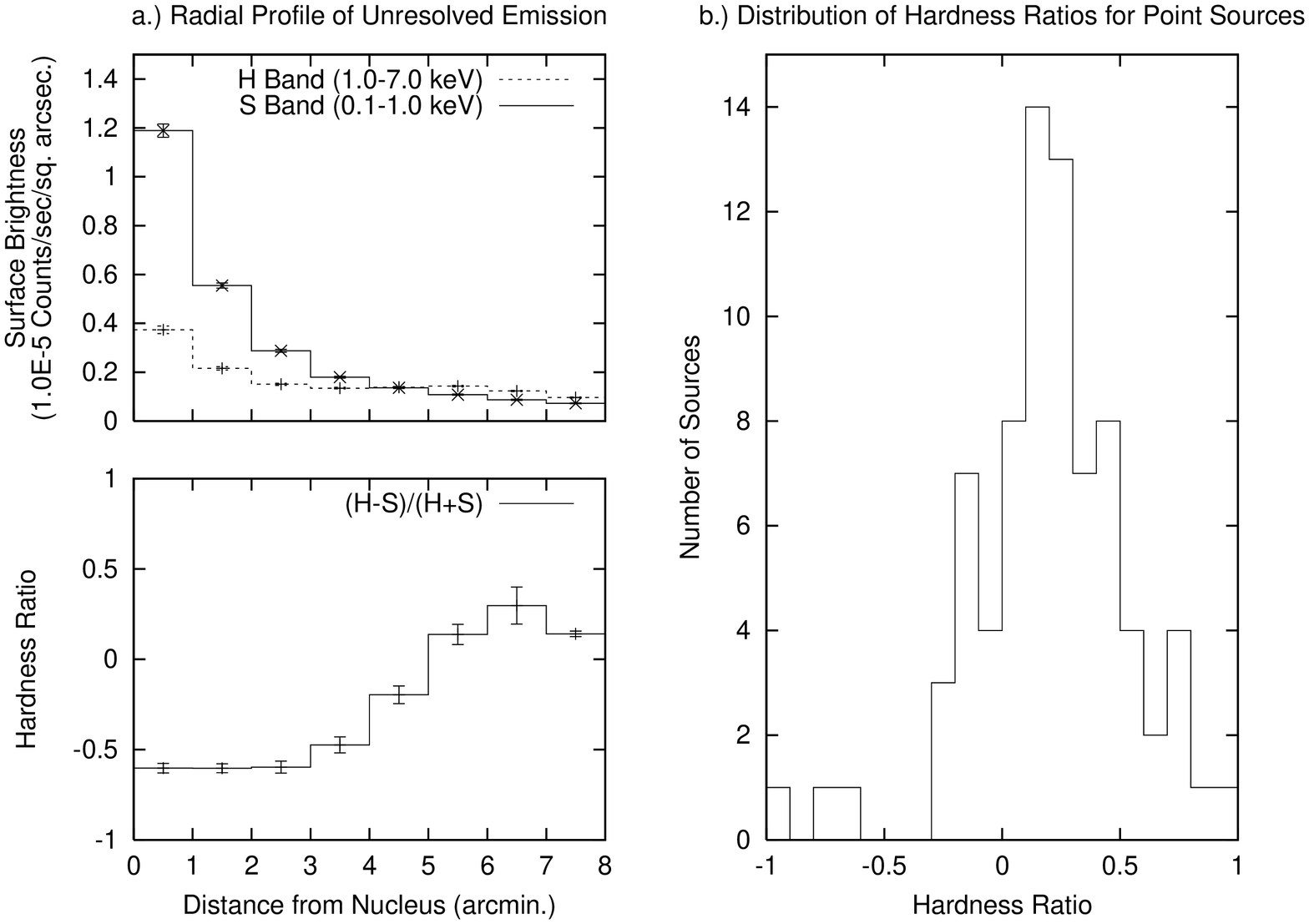,width=130truemm,angle=-90.0}}
\protect\caption{ a.) Radial Profile of Unresolved Emission. The hardness ratio is
computed using net counts, with background taken from the outermost
bin.
b.) Distribution of Hardness Ratios for Point Sources, from Garcia et al. 2000}
\end{figure}

With respect to
the related question
of the contribution of LMXBs with soft spectral components,
Chandra's
excellent spatial resolution allows us to
study the x-ray
spectra of individual bright
point sources without the source confusion
problems that complicate the analysis of ROSAT PSPC or ASCA
data. Moreover, background is negligible ($<1$ count) in these
observations, so we avoid additional uncertainties due to 
contamination by background sources. At present, we have analysed the spectra
of eight bright sources within $5^{\prime}$ of the nucleus. To
simplify our analysis, we have limited the data to a single 6.4 ksec ACIS-I
observation, so that only one response matrix for each source was
required. 
Our results are shown in Table 1. All spectra are
adequately fit with a simple bremsstrahlung model, but we do not have
sufficient data to exclude a soft spectral
component. The unusually hard spectrum for CXO004238.5+411604 
may be the result of CCD pile-up, in which two or more x-rays
counted in the same CCD pixel within one readout frame are mistaken as a
single higher-energy x-ray. It is interesting to note that the
temperatures appear to decrease with intensity, to relatively soft
($\sim$3 keV) values for the faintest sources.
Clearly,
more data and more careful analysis are required.
\begin{table}[htb]
\caption{Spectral Fits to Individual Sources}
\begin{center}
\begin{tabular}{|c|c|c|c|c|}
\hline 
Source&
Counts&
kT (keV)&
N\( _{H} \) (\( 10^{21} \) atoms cm\( ^{-2} \) )&
\( \chi ^{2}/DOF \)\\
\hline 
\hline 
004238.5+411604&
1260&
200 (!)&
2.6\( \pm  \)0.3&
152/115\\
004248.4+411522&
670&
7.9\( \pm 2.3 \)&
1.0\( \pm 0.3 \)&
53/59\\
004242.0+411609&
600&
9.6\( \pm  \)3.4&
1.7\( \pm 0.4 \)&
52/53\\
004254.8+411603&
525&
5.3\( \pm  \)1.4&
1.0\( \pm 0.3 \)&
57/47\\
004231.0+411623&
245&
4.3\( \pm  \)1.4&
0.5\( \pm  \)0.4&
27/22\\
004247.0+411629&
200&
2.4\( \pm  \)0.9&
1.3\( \pm 0.6 \)&
7/16\\
004246.8+411616&
200&
2.7\( \pm 0.8 \)&
1.9\( \pm 0.7 \)&
18/16\\
004247.7+411533&
140&
2.6\( \pm 1.1 \)&
\( <0.7 \)&
5/10\\
\hline 
\end{tabular}
\end{center}
\end{table}

\section{Probing the ISM Toward Individual X-ray Sources}  

In some of the sources listed in Table 1, there appears to be
a significant column density in excess of the galactic value of
$\sim7\times10^{20}$~atoms~cm$^{-2}$. This is likely to contain
contributions both from the ISM and from absorption at the
source. As has been pointed out during this workshop, the modest
range of $N_H$ in Table 1 would be unexpected from the ISM, which
tends to be very clumpy. Given more data and a better understanding
of the intrinsic absorption, it may be possible
to use x-ray spectra as independent measures of extinction due to
the ISM. More sophisticated models, in which galactic and
extragalactic extinction (using different metallicities) are treated
separately, will be required.

The ROSAT PSPC Survey (Supper
et al. 1997) reports 15 counterparts to optically identified SNRs,
although several are ambiguous.  A deep ROSAT HRI survey (Magnier et
al. 1997) identifies six candidates brighter than
$\sim8\times10^{35}$~ergs~s$^{-1}$, although again, half of these have
other possible counterparts. We have compared the sources
detected during standard HRC data processing with optically identified
SNRs from Braun \& Walterbos (1993) and Magnier et al. (1995) and find
no new candidates. However,
our current sensitivity does not exceed that in the HRI survey.
At the end of a year's time we should achieve
comparable sensitivity, particularly for the more compact remnants.
Given the poor correlation between optical and x-ray emission
of SNRs in M31 and M33 (Duric, these proceedings), however, we may need to wait
for radio catalogs of SNRs to improve our identification efficiency.

\section{Conclusions}

Clearly, we have only just begun our study of the ISM with Chandra,
and the standard caveats which accompany preliminary analyses from new
instruments apply here. However, the results to date suggest that we
will be able in time to significantly improve our understanding of the
x-ray properties of M31's ISM.

\section*{Acknowledgments}
This work was supported in part by NASA Grant NAG5-8358 and NASA
Contract NAS8-39073. 

\section*{References}\noindent

\references

Braun, R. \& Walterbos, R. \Journal{1993}{\em Astron. Astrophys.,
Suppl. Ser.}{98}{327}

Borozdin, K. \& Priedhorsky, W. (2000) Astro-ph 0006119

Garcia, M., Murray, S.S., Primini, F.A., Forman, W.R., McClintock,
J.E., \& Jones, C. (2000){\ApJL} in press

Irwin, J.A. \& Bregman, J.N. \Journal{1999}{\ApJ}{527}{125}

Magnier, E.A., Primini, F.A., Prins, S., van Paradijs, J., \& Lewin,
W.H.G. \Journal{1997}{\ApJ}{490}{649}

Magnier, E.A., Prins, S., van Paradijs, J., Lewin, W.H.G., Supper, R.,
Hasinger, G., Pietsch, W., \& Trumper, J. \Journal{1995}{\em
Astron. Astrophys., Suppl. Ser.}{114}{215}

Primini, F.A., Forman, W., \& Jones, C. \Journal{1993}{\ApJ}{410}{615}

Supper, R., Hasinger, G., Pietsch, W., Trumper, J., Jain, A., Magnier,
E.A., Lewin, W.H.G., \& van Paradijs,
J. \Journal{1997}{\AAp}{317}{328}

West, R.G., Barber, C.R., \& Folgheraiter, E.L. \Journal{1997}{\MNRAS}{287}{10}
%
%
%
%
%
%

\end{document}